\documentclass[aps,prb,twocolumn,final,groupedaddress]{revtex4-1}
\pdfoutput=1

\usepackage[intlimits]{amsmath}
\usepackage{amssymb}
\usepackage{graphicx}
\graphicspath{{./Figures/}}
\usepackage{braket}
\usepackage{color}
\definecolor{MidnightBlue}{RGB}{0,0,160}
\usepackage[pdftitle={Properties of a-Si:H from ab initio}, pdfauthor={Philippe Czaja}, colorlinks=true, citecolor=MidnightBlue,linkcolor=black,urlcolor=MidnightBlue, pdftex, breaklinks=true]{hyperref}
\usepackage[caption=false]{subfig}
\usepackage{floatrow}
\usepackage{booktabs}

\newcommand{\ve}[1]{\mathbf{#1}}
\newcommand{\un}[1]{\text{ #1}}

\newcommand{\md}[1]{\mathrm d #1 }

\newcommand{\mt}[1]{\text{ #1}}

\newcommand{\abs}[1]{\left\lvert #1 \right\rvert}
\newcommand{\figref}[1]{FIG. \ref{#1}}

\begin{document}

\title{Ab-initio analysis of structural, electronic, and optical properties of a-Si:H}

\author{Philippe Czaja}
\email{p.czaja@fz-juelich.de}
\author{Urs Aeberhard}
\affiliation{IEK-5 Photovoltaik, Forschungszentrum J\"ulich, 52425 J\"ulich, Germany}

\author{Massimo Celino}
\author{Simone Giusepponi}
\affiliation{ENEA, C. R. Casaccia, Via Anguillarese 301, 00123 Rome, Italy}

\author{Michele Gusso}
\affiliation{ENEA, C.R. Brindisi, 72100 Brindisi, Italy}

\date{\today}

\begin{abstract}
We present a first-principles study of the structural, electronic, and optical properties of hydrogenated amorphous silicon (a-Si:H).
To this end, atomic configurations of a-Si:H with 72 and 576 atoms respectively are generated using molecular dynamics. Density functional theory calculations are then applied to these configurations to obtain the electronic wave functions. These are analyzed and characterized with respect to their localization and their contribution to the density of states, and are used for calculating ab-initio absorption spectra of a-Si:H. The results show that both the size and the defect structure of the configurations modify the electronic and optical properties and in particular the value of the band gap. This value could be improved by calculating quasi-particle (QP) corrections to the single-particle spectra using the G$_0$W$_0$ method. We find that the QP corrections can be described by a set of scissors shift parameters, which can also be used in calculations of larger structures.
\end{abstract}

\keywords{hydrogenated amorphous silicon, molecular dynamics, electronic structure, optical properties}

\pacs{}

\maketitle

\section{Introduction}
Hydrogenated amorphous silicon (a-Si:H) has been used as a cheap and efficient absorber material in silicon thin-film solar cells for more than 40 years \cite{carlson:76}, and has lately found another application in photovoltaics as a passivation layer in silicon-heterojunction cells. Understanding its microscopic  structure in order to optimize its macroscopic properties for the application in photovoltaics has motivated several ab-initio studies of a-Si:H throughout the years \cite{allan:82,fedders:93,tuttle:98,valladares:01,jarolimek:09,curioni:11,legesse:13}. Two principle challenges are thereby met. First, a model atomic structure has to be generated that correctly reproduces certain experimental features of a-Si:H, such as the defect density, the radial pair correlation function, or the vibrational properties. Second, the electronic structure has to be calculated on a level that allows the extraction of physically meaningful macroscopic properties. From the viewpoint of photovoltaics, special interest lies on the description of the optical properties and on the identification and characterization of localized defect states, which have a crucial impact on the device performance due to their role as recombination centers in non-radiative recombination \cite{Shockley1952}.

Whereas the generation of defective a-Si:H configurations, i.e., configurations containing dangling bonds, is instructive for studying the origin and the nature of localized defect states, these configurations are not well suited for obtaining realistic macroscopic properties, due to an overestimation of the defect density. In fact, structures containing one defect need to have a size of at least $10^6$ atoms to yield realistic defect densities \cite{street:91}, which is out of the range of current studies dealing with structure sizes of the order of 1000 atoms. The generation of defect-free configurations is therefore an important step towards a full ab-initio description of a-Si:H. However, for a long time defect-free configurations of a-Si and a-Si:H could be generated only with model approaches 
such as the Wooten-Winer-Weaire algorithm \cite{drabold:00}, the Bethe-lattice approach \cite{allan:82}
or the Reverse Monte-Carlo approach \cite{celino:99}.
Only recently, large-scale ($\sim$500 atoms) atomistic simulations of a-Si:H using a quench-from-a-melt approach \cite{curioni:11,nolan:12} combining both classical and ab-initio molecular dynamics (MD) have been reported to yield configurations of low defect density \cite{legesse:13}. 
The same approach was used to generate low-defect and even defect-free a-Si:H configurations of 72 atoms 
within ab-initio MD \cite{jarolimek:09}.
 Following this approach, we present atomistic ab-initio MD simulations on a-Si:H, resulting in defective 
and defect-free configurations of 72 and 576 atoms respectively.

The calculation and analysis of the electronic structure and the optical properties of a-Si:H on the Density-Functional-Theory (DFT) level has been the subject of multiple recent works \cite{tuttle:98,valladares:01,jarolimek:09,curioni:11,legesse:13}. The focus of interest in these works has been mainly on the origin of mid-gap states and band tails, and on the effect of hydrogen concentration and structural features on the mobility gap and the optical gap respectively. Very little attention has however been paid to the effect of computational artifacts on the electronic and optical properties. In particular, two effects should be taken into account when trying to reproduce the experimental properties of a-Si:H, and are therefore investigated in this work: the effect of the super cell size and the effect of many-body interactions. A recent work stated that finite size effects do not play any role for structures larger than 72 atoms \cite{jarolimek:09}, which however disagrees with our findings. The incomplete description of many-body effects on the other hand is a well-known problem of standard DFT \cite{perdew:85}, and is the reason why the optical and mobility gaps are severely underestimated in previous studies using the local density approximation (LDA) or the generalized gradient approximation (GGA). Good values for the gaps have however been achieved recently with hybrid functionals \cite{legesse:13}. In this work we try to incorporate many-body interactions systematically by explicitly calculating the quasi-particle corrections \cite{hybertsen:85} to the Kohn-Sham energies. These corrections are often described by a heuristic approach -- termed \emph{scissors shift} (SS) \cite{godby:88} -- , where the electron energies are simply shifted to fit the experimental band gap. Since a distinct experimental value of the band gap of a-Si:H does however not exist, a set of shifting parameters can only be determined from a GW calculation. Here we present the results of such a calculation.

\section{Technical details}
Born-Oppenheimer molecular dynamics (BOMD) simulations and electronic structure calculations are performed on the DFT \cite{hohenberg:64,kohn:65} level, using a PBE-GGA exchange-correlation functional \cite{pbe} and periodic boundary conditions (PBC), meant to mimic an infinitely extended system.

For the BOMD simulations of the small structure (72 atoms) the PWscf (Plane-Wave Self-Consistent Field) code of the Quantum ESPRESSO suite \cite{giannozzi:09,qe} is used with ultrasoft pseudopotentials, whereas for the large structure (576 atoms) the Quickstep code of the CP2K suite \cite{cp2k} is used with norm-conserving Goedecker-Tetter-Hutter pseudopotentials \cite{goedecker:96, hartwigsen:98, krack:05}. All MD simulations are restricted to the $\Gamma$-point.

The electronic structure is calculated with the PWscf code of the Quantum ESPRESSO package using norm-conserving pseudopotentials. $\ve{k}$-point summations are carried out on a $4\times4\times4$ grid for the small system, and on a $2\times2\times2$ grid for the large system. The plane wave cut-off energy is set to 52 Ry. These parameters were chosen by checking the convergence of the total energy of the system.

Quasi-particle corrections to the Kohn-Sham energies for the small configuration are obtained by performing single-shot G$_0$W$_0$ calculations \cite{Hedin1965} with the BerkeleyGW code \cite{Deslippe2012} within the generalized plasmon-pole (GPP) approximation \cite{lundqvist:67,overhauser:71,hybertsen:86} on a $2\times2\times2$ grid using 3000 bands and a kinetic energy cut-off of 10 Ry. These values were chosen by checking the convergence of the LUMO-HOMO gap with respect to all three parameters simultaneously. The Kohn-Sham wave functions are retained as they are assumed to differ very little from the quasi-particle wave functions \cite{hybertsen:86}. The GPP approximation is chosen because it has the advantage of requiring the dielectric tensor $\epsilon(\omega)$ only in the static limit $\omega\to 0$, while still taking into account the effects of dynamical screening. This ensures an accuracy similar to a full-frequency calculation for many semiconductors, including c-Si \cite{larson:13, hybertsen:86}.

The BerkeleyGW code is also used for calculating the optical properties within linear-response theory using the random phase approximation (RPA) \cite{ehrenreich:65}. Electron-hole interaction is disregarded as it is generally assumed to have no significant effect on the absorption spectra of amorphous semiconductors \cite{street:91}.

\section{Generation of the atomic structure}
The first step towards the ab-initio description of a-Si:H is the generation of physically meaningful atomic configurations, i.e., configurations that reproduce the experimental properties of a-Si:H. Our method of choice to achieve this is a simulated annealing quench-from-a-melt protocol \cite{car:88,curioni:11,nolan:12}. With this method two kinds of configurations are generated, one consisting of 64 Si + 8 H atoms, which we will refer to as the small system, and one consisting of 512 Si + 64 H atoms, which we will refer to as the large system. Both structures are cubic with a size of $a=11.06$ \AA\ and $a=22.12$ \AA\ respectively, resulting in a density of $2.214$ g/cm$^3$ that matches the experimental value \cite{curioni:11}. A hydrogen concentration of about 11\% is chosen as this is the nominal concentration set in experimental materials optimized for photovoltaic performance \cite{johlin:13}.

\subsection{Small system}
\begin{figure}
\includegraphics[width=0.3\textwidth]{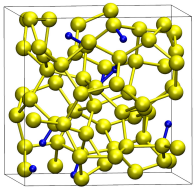}
\caption{Snapshot of the small a-Si:H configuration in the simulation box. Hydrogen atoms and bonds with Silicon atoms are blue, Silicon atoms and their bonds are yellow.}
\label{amorfo72_finale} 
\end{figure}

The small structure is generated by randomly inserting H atoms in a crystalline Si cell,
avoiding placing atoms too near to each other to minimize the energy in the system. A long (more than 50 ps) BOMD simulation is then performed on the system at constant volume and temperature 
$T$= 2000 K, controlled by an Andersen thermostat \cite{andersen:80}.
This simulation assures that any remaining crystalline symmetry in the atomic configuration is destroyed.
During the high temperature simulation, atoms cover a distance of about 2 nm, which ensures that 
the final configuration retains no memory of the initial geometry.
Afterwards the temperature is lowered to $T$= 1200 K within 5 ps and held constant for another 5
ps, as an intermediate step prior to the final quench to $T$ = 300 K. This takes another 5 ps, which results in a high quench rate of $\sim 10^{14}$ K/s.
The quenched configuration is then used as a starting point for a final BOMD simulation at room temperature to 
fully thermalize the amorphous system.
After 30 ps of simulation another 5 ps are used to compute average physical quantities and to characterize the system with respect to structural and electronic properties. The final configuration at the end of the simulation is shown in \figref{amorfo72_finale}.

\subsection{Large system}
\begin{figure}
\begin{center}
\includegraphics[width=0.6\textwidth]{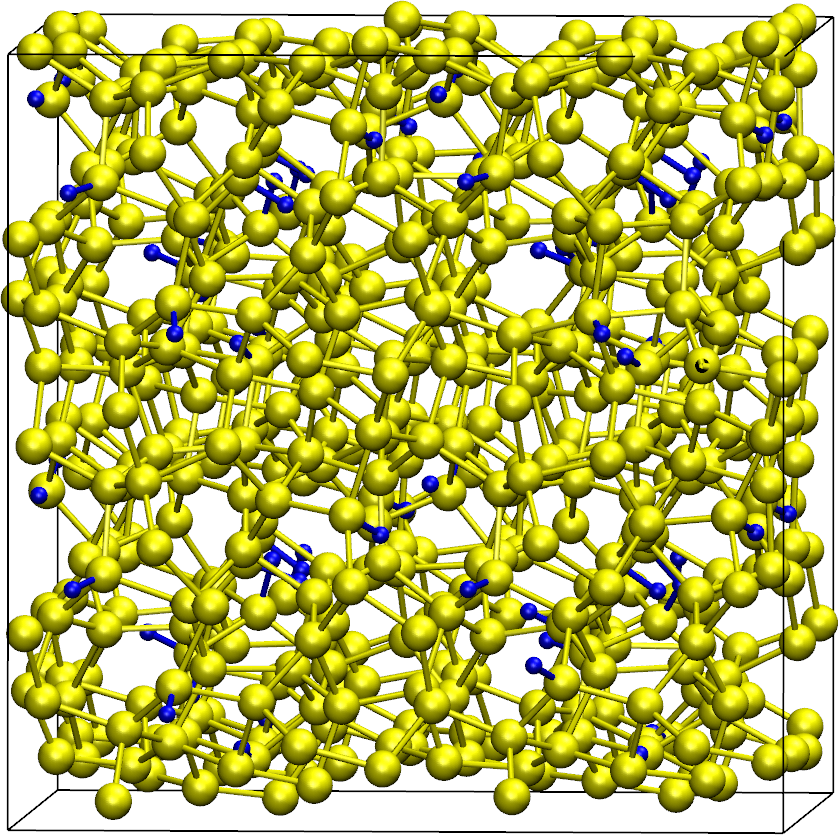}
\caption{Snapshot of the large a-Si:H configuration in the simulation box. Hydrogen atoms and bonds with Silicon atoms are blue, Silicon atoms and their bonds are yellow.}
\label{amorfo576_finale} 
\end{center}
\end{figure}

The starting configuration for the large system is produced by replicating the small system in all directions. Since a high temperature annealing was already performed for the small system, a low temperature annealing in order to minimize the spurious defects at the internal interfaces is sufficient.
For that purpose BOMD simulations are performed for 80 ps in time steps of 20 a.u. at constant volume. The temperature is controlled by a Nos\'e thermostat \cite{nose:84}. Within the first 60 ps the temperature is modified from 300 K to 600 K and back to 300 K in steps of
100 K.
Afterwards, an additional simulation run at $T$= 300 K is performed for 20 ps.
\figref{amorfo576_finale} shows the final configuration at the end of the simulations.

\section{Structural properties}
\begin{figure}
\begin{center}
\includegraphics[width=\textwidth]{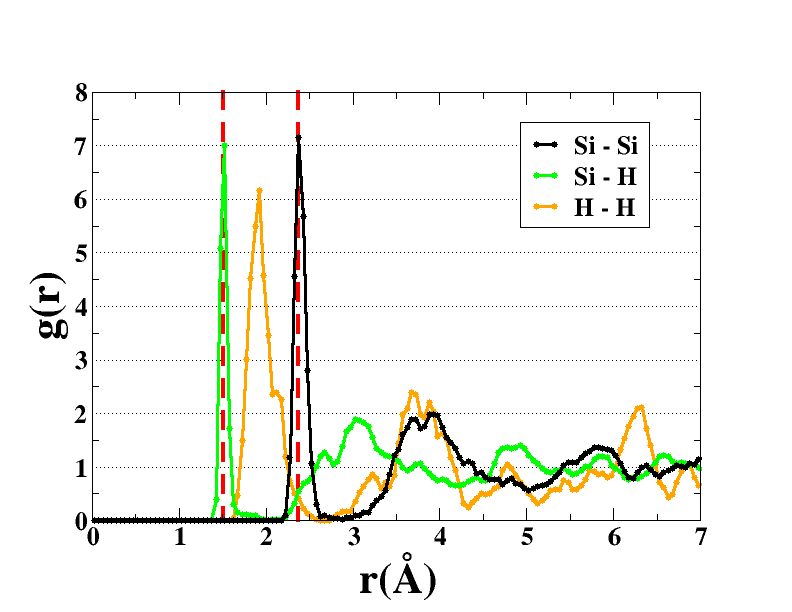}
\includegraphics[width=\textwidth]{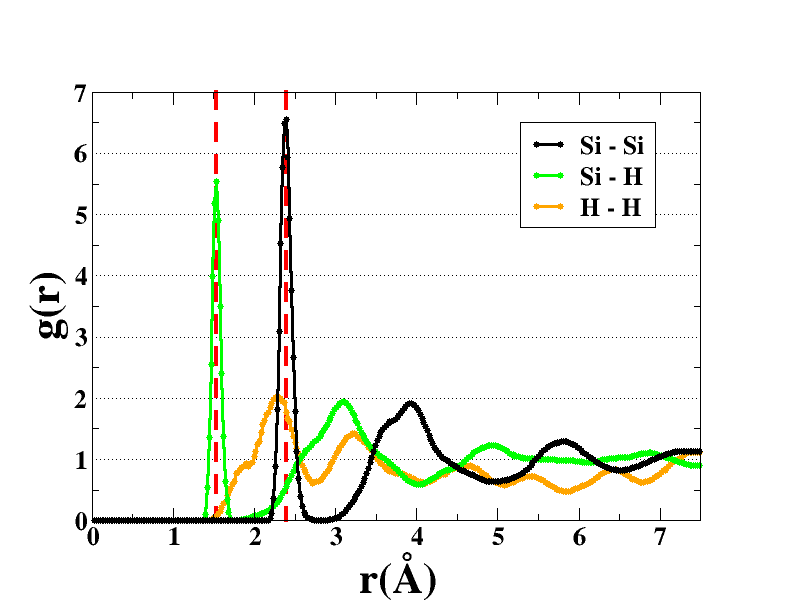}
\caption{Radial pair correlation functions $g(r)$ of the small system (top) and the large system (bottom) for Si-Si pairs (black), Si-H pairs (green), and H-H pairs (orange), averaged over the last 5 ps of the MD simulation.}
\label{gdr} 
\end{center}
\end{figure}

In order to characterize the system and estimate its quality in terms of how it compares to real a-Si:H we study the radial pair correlation function $g(r)$, defined as \cite{jarolimek:09}
\begin{equation}
g_{\alpha\beta}(r)=\frac{V}{4\pi r^2 N_{\alpha}N_{\beta}}\sum_{I=1}^{N_{\alpha}}\sum_{J=1}^{N_{\beta}}\delta(r-\abs{\ve R_I - \ve R_J}) \mt{.}
\end{equation}
The characterization starts after the system has reached thermal equilibrium. \figref{gdr} shows $g(r)$ for both systems averaged over the last 5 ps of the simulation, revealing sharp peaks at $2.37$ \AA\ for Si-Si pairs, and at $1.51$ \AA\ for Si-H pairs.

\begin{figure}
\begin{center}
\includegraphics[width=1.\textwidth]{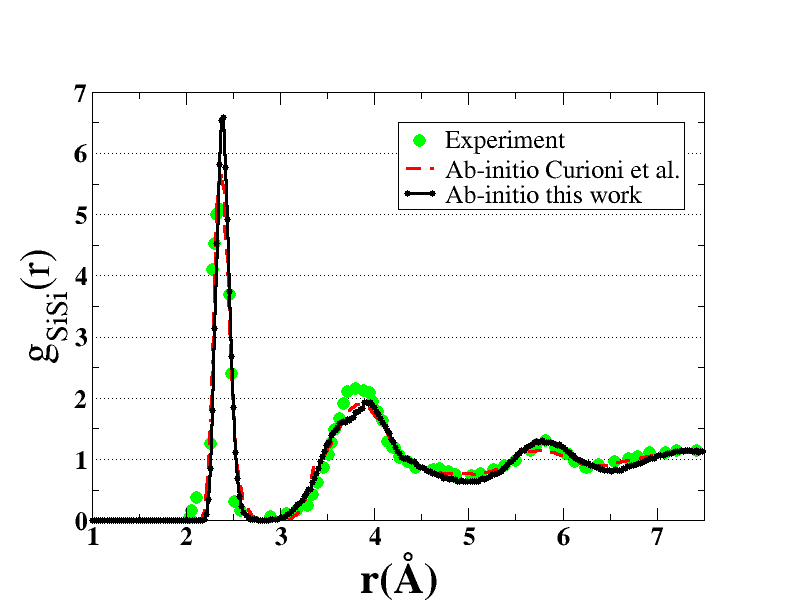}
\caption{Comparison between experimental and computed (large system) Si-Si radial pair correlation functions $g_{Si-Si}(r)$. 
The experimental data is taken from Ref. \onlinecite{bellisent:89}. The Curioni curve is reported in Ref. \onlinecite{curioni:11}.
}
\label{amorfo576_gdr_sisi} 
\end{center}
\end{figure}

The comparison of $g_{Si-Si}(r)$ for the large system with another recent work by Curioni et al. \cite{curioni:11} as well as experimental data for a-Si:H \cite{bellisent:89} (\figref{amorfo576_gdr_sisi}) shows very good agreement at all distances, indicating that our configurations well reproduce the properties of real a-Si:H (even though the experimental results are obtained from non-hydrogenated amorphous Silicon). 
Analyzing the data, we observe that our curve has a slightly higher first peak which indicates that our system has a higher 
percentage of four-fold coordinated atoms. 
This is more evident in the comparison with the experimental first peak, because a-Si has even more unsaturated Si atoms.

To gain a deeper understanding of the structural properties, a coordination analysis of the Si atoms is performed
and reported in Table~\ref{coord1}. On the basis of the calculated radial pair correlation functions, a geometrical criterion
is used to identify the nearest neighbors in the coordination analysis, applying a distance cutoff of 2.85 \AA\ and 1.7 \AA\, for Si-Si and Si-H pairs, respectively. 
These values are used to characterize both systems. Concerning the small system, it is observed that the average number of
neighbors of a Si atom is 3.96. In details, 4 Si atoms have three-fold coordination (6.3\%), 
58 Si atoms have four-fold coordination (90.6\%) and the remaining 2 Si atoms have five-fold coordination (3.1\%). 
In contrast, the large system has an average coordination number equal to 3.99 because 99.0\% of the Si atoms have fourfold
coordination. Only 4 Si atoms (0.8\%) have threefold coordination and 1 Si atom is five-fold coordinated (0.2\%).
In conclusion, the large systems is more ordered with a higher percentage of four-fold coordinated Si atoms.
This trend is confirmed by a more precise analysis of the bonds based on the computation of the Electron Localization Function (ELF)
\cite{becke:90}. This method will be explained in the next section and the discrepancies with the geometric approach will be discussed.

To understand which type of bonding is formed around every Si atom and which role the H atoms play, the environment of each Si
atom is analyzed. To this end, we compute the number of Si$x$H$y$ configurations, where $x$ and $y$ are the
number of neighboring Si and H atoms of the considered Si atom, respectively. The results are reported in Table~\ref{coord2}, together with a comparison between the results obtained from the geometrical and the ELF method.
In both systems it is clear that no H$_2$ dimers are formed and no two H atoms are bonded to the same Si atoms, 
since all the Si$x$H2 environments are equal to zero.
On the contrary, almost all the H atoms are saturating dangling bonds of Si atoms to form four-fold coordinated Si atoms.
Indeed, for example in the large system, all the 64 H atoms are involved in Si3H1 environments. 
This analysis confirms that the large system is more ordered than the small one.

\begin{table}
\caption{Coordination analysis for Si atoms of the a-Si:H systems computed in two ways: the geometric protocol and the ELF approach.
The number Si atoms (with the percentage in parenthesis) with given coordination (Coord) is shown.}
\begin{tabular}{@{}c|cc|cc@{}}
\hline
\hline
%	&\multicolumn{4}{c}{$n_t (\%)$}	\\
Coord&\multicolumn{2}{c|}{Small system}&\multicolumn{2}{c}{Large system}\\
	& Geom.	& ELF 	& Geom.	& ELF	\\ 
\hline 
1	& 0		& 0		& 0		& 0		\\
2	& 0		& 1 (1.6)	& 0		& 0		\\
3	& 4 (6.3)	& 0		& 4 (0.8)	& 0 		\\
4	& 58 (90.6)	& 63 (98.4)	& 507 (99.0)	& 512 (100) \\
5	& 2 (3.1)	& 0 		& 1 (0.2)	& 0 		\\
\hline
\hline
\end{tabular}
\label{coord1}
\end{table}

\begin{table}
\caption{Environment analysis for Si atoms of the a-Si:H systems computed in two ways: the geometric protocol and the ELF approach.
For both systems the following is reported: the coordination  (Coord), the type of environment (Envir),
the number of Si atoms with that Envir.
}
\begin{tabular}{@{}cc|cc|cc@{}}
\hline
\hline
%		&\multicolumn{4}{c}{$n_e $}	\\
Coord & Envir	&\multicolumn{2}{c|}{Small system}& \multicolumn{2}{c}{Large system}\\
	&	& Geom.	& ELF	& Geom.	& ELF \\ 
\hline 
1	&Si1H0	& 0		& 0		& 0		& 0	\\
	&Si0H1	& 0		& 0		& 0		& 0	\\
\hline 
	&Si2H0 	& 0 		& {\bf 1}	& 0		& 0 	\\
2	&Si1H1	& 0		& 0		&  0		& 0	\\
	&Si0H2	& 0		& 0		&  0		& 0	\\
\hline 
	&Si3H0	& {\bf 4} 	& 0		&  {\bf 4} 	& 0	\\
3	&Si2H1	& 0		& 0		&  0		& 0	\\
	&Si1H2	& 0		& 0		&  0		& 0	\\
\hline 
	&Si4H0	& {\bf  51}	& {\bf 55}	& {\bf 443}	& {\bf 448}	\\
4	&Si3H1	& {\bf 7}	& {\bf 8}	& {\bf 64}	& {\bf 64}	\\
	&Si2H2	&0		& 0		&  0		& 0	\\
\hline 
	&Si5H0	& {\bf 1}	& 0		& {\bf  1}	& 0	\\
5	&Si4H1	& {\bf 1}	& 0		&  0		& 0	\\
	&Si3H2	& 0		& 0		& 0		& 0	\\
\hline
\hline
\end{tabular}
\label{coord2}
\end{table}

\section{Electronic properties}

\subsection{Bonding}

\begin{figure}
\begin{center}
\includegraphics[width=1.\textwidth]{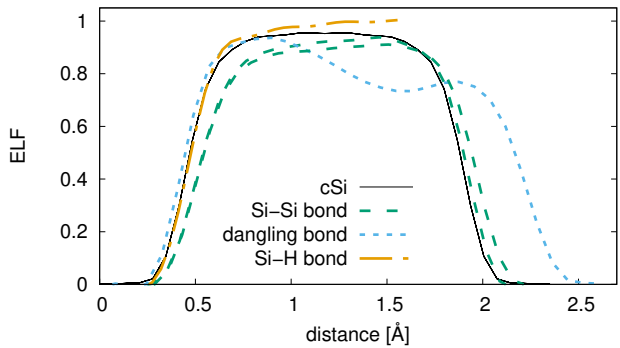}
\caption{ELF along the bonding axes for a three-fold bonded Si atom in a-Si:H. One can distinguish two Si-Si bonds (dashed), one dangling bond (dotted), and one Si-H bond (dash-dotted). The ELF in c-Si is shown as a reference (solid).}
\label{elf} 
\end{center}
\end{figure}

In order to relate the electronic properties to the atomic structure, and in particular identify structural defects, we use the electron localization function (ELF) \cite{becke:90}, which enables us to determine the coordination of each atom by identifying covalent and dangling bonds. A covalent bond is formed by overlapping atomic orbitals, which results in an accumulation of charge between the bonded atoms. This accumulation shows as a broad maximum in the ELF along the bonding axis for non-polar covalent semiconductors such as c-Si \cite{savin:92}, for which it reaches a value of 0.95 (\figref{elf}). As the bond breaks, charge is no longer localized between the atoms, but in atomic orbitals instead, which shows as peaks in the ELF near the atomic positions, forming a minimum in the center (\figref{elf}). This difference in the behavior of the ELF along the axes between neighboring atoms can be used to distinguish whether a bond exists or not. As the different shape also goes along with a distinctly different value of the ELF in the center between the atoms, we can alternatively use this value as a simple criterion for identifying a dangling bond. Even though the values of the maxima and minima, respectively, vary in amorphous semiconductors, a value of 0.8 has been found to separate the maxima from the minima for all analyzed Si-Si and Si-H bonds in a-Si:H, and is therefore used as a threshold throughout this study. This threshold can also be applied to Si-H bonds, for which the ELF is also shown in \figref{elf}. The reason why the ELF drops to zero near the Si but not near the H atoms is that the core electrons of Si are not taken into account.

By systematically calculating the ELF along the bonding axis between all pairs of neighboring atoms and applying the criterion described above, we find that the small configuration contains exactly one Si atom that is only two-fold coordinated, whereas all other Si atoms are perfectly four-fold coordinated, and all the H atoms are bonded to exactly one Si atom (see Table~\ref{coord1}). This translates to a defect density of $1.5\times10^{21} \un{cm}^{-3}$, which is about five orders of magnitude higher than experimentally measured values \cite{favre:87, street:89}. The large configuration does not contain any dangling bonds, that is, all Si atoms are four-fold and all H atoms are one-fold coordinated (see Table~\ref{coord1} and Table~\ref{coord2}). The discrepancy between the ELF and the geometrical approach can be explained by the presence of strain. 
We have checked that the number of compressed and elongated bonds are on average the same and less than 2\%.
The strain is small but not negligible and it can induce errors in the computation of the coordination.
Thus the ELF approach, with its ability to ascertain the presence of covalent bonds, predicts that the large system is a defect-free a-Si:H configuration and is hence more promising from the viewpoint of reproducing the electronic 
properties of the real amorphous material.

\subsection{Energy and localization of wave functions}

We investigate the effect of the structure and the size of the configurations on the electronic properties in terms of the density and localization of states. 
As a quantitative measure for the localization of a wave function $\psi$ we use the spread $S$, which is calculated as the square root of the variance of $|\psi|^2$ with respect to one super cell:
\begin{align}
  S &= \sqrt{4\left(\braket{\ve r^2} - \braket{\ve r}^2\right)} \\ &= 2\cdot\sqrt{\int_{\Omega}\md{\ve r} |\psi(\ve r)|^2 \ve r^2 - \left(\int_{\Omega} \md{\ve r} |\psi(\ve r)|^2 \ve r \right)^2} \text{ ,}
\end{align}
where we assume that $\psi$ is normalized with respect to the super cell volume $\Omega$. It can be easily seen that a maximally localized $\psi$ (i.e., a Dirac delta function) gives $S = 0$, whereas a wave function that is maximally delocalized over the super cell (i.e., a plane wave) will result in $S = a$, where $a$ is the lattice constant of the super cell. This explains the factor $2$ in the definition.
As $S$ is not uniquely defined but will in general depend on the choice of the integration volume $\Omega$, we use the \emph{minimal spread}
\begin{equation}
S_{min} = \min_\Omega S
\end{equation}
instead. Throughout the rest of the paper the term \emph{spread}, or $S$ respectively, will therefore refer to $S_{min}$. Using the integration volume $\Omega_m$ that minimizes the spread we can also define a wave function center
\begin{equation}
\braket{\ve r} = \int_{\Omega_m} \md{\ve r} |\psi(\ve r)|^2 \ve r \text{ ,}
\end{equation}
which can be interpreted as the position where the wave function is localized.
This definition allows us to identify localized states, and to locate them both in real and in energy space.

\begin{figure}
\begin{center}
\includegraphics[width=1.\textwidth]{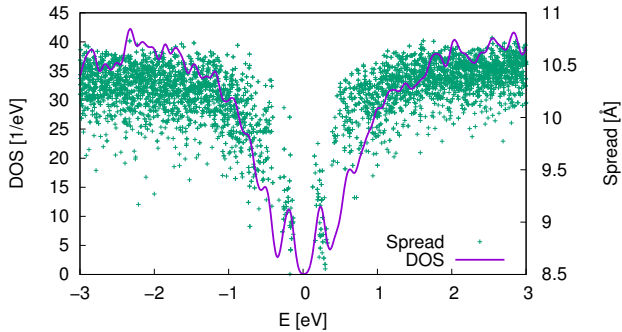}
\caption{DOS and spread of the small system. Each dot represents one wave function.}
\label{spread_64} 
\end{center}
\end{figure}

\begin{figure}
\begin{center}
\includegraphics[width=1.0\textwidth]{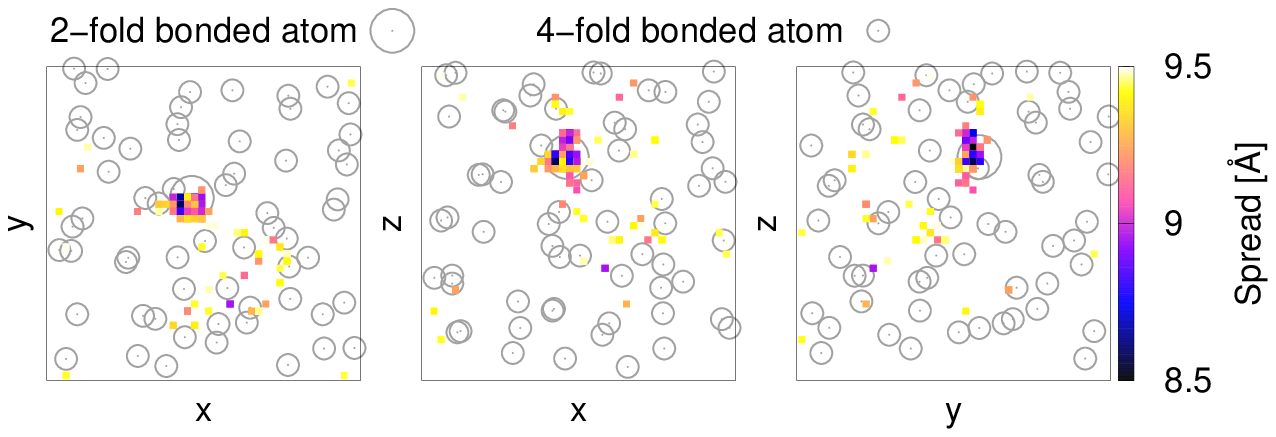}
\includegraphics[width=1.0\textwidth]{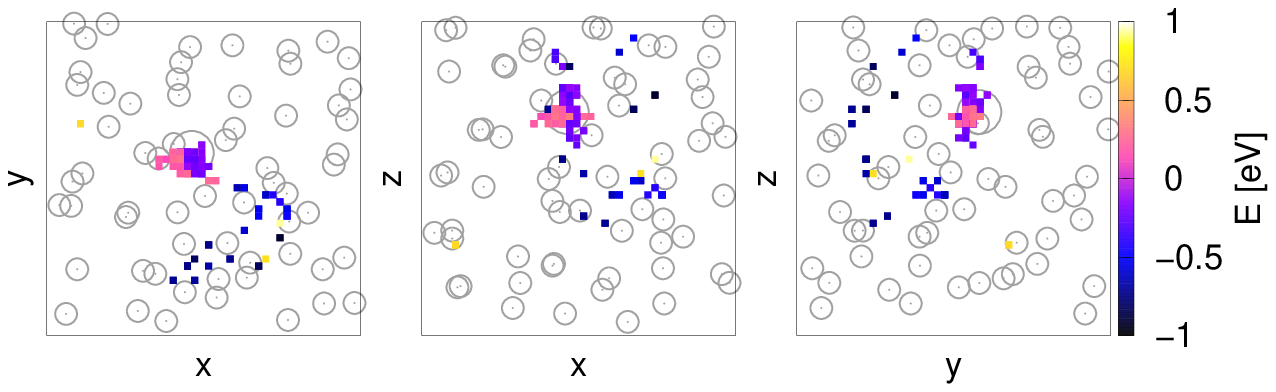}
\caption{Distribution of localized states with $S\le 9.5 \un{\AA}$ in the super cell (projections onto the $xy$-, $xz$-, and $yz$-plane). Each dot represents one wave function, where the colors denote its spread (top) and its energy (bottom) respectively. The circles represent the Si atoms, where different sizes stand for different coordination.}
\label{3d_64} 
\end{center}
\end{figure}

In \figref{spread_64} the spread is shown as a function of the energy together with the DOS for the small configuration. The DOS reveals two distinct peaks inside the gap that originate from localized states, as the spread shows. 
In order to understand the origin of these states we use the definition of the wave function center to determine the position of localized states up to a spread of $9.5\un{\AA}$ inside the super cell (\figref{3d_64}). Each dot represents one wave function, where the color represents the spread (top) and the energy (bottom) of that wave function. Additionally the position of the atoms are symbolized by red dots, where the larger dot represents the single two-fold bonded atom. The map shows that the strongly localized states inside the gap are localized directly at the under-coordinated atom, and can therefore be assigned to dangling bonds. Moreover, the bottom figure shows that the states below and above the Fermi energy (0 eV) are spatially separated and thus originate from different dangling bonds.

Whereas the gap states can be easily distinguished, the distinction of the tail states in \figref{spread_64} is not so trivial. Even though a tail region with semi-localized states and a band region with mostly extended states can be identified, the relatively small difference in the spread makes the accurate determination of the mobility edges, i.e., the transitions between regions of different localization, very difficult. The way this is usually done is by setting a threshold for the localization that separates localized and extended states \cite{legesse:13}. This gives an estimate for the mobility gap, which is defined as the gap between extended valence and conduction band states \cite{winer:91}. The threshold for the spread should be set such that band states far away from the gap exceed it. Clearly this method becomes more accurate the more unambiguous the choice of the threshold is, and the steeper the localization changes at the mobility edges. By choosing a value of $S=10.1\un{\AA}$ we obtain a mobility gap of roughly 0.9 eV for the small configuration.

The difficulty in distinguishing band and tail states is to some extent an artifact of the finite super cell size. This can be understood by considering that even the band states are not perfectly delocalized plane-waves but have regions of higher and lower probability density, leading to a spread smaller than the extension of the super cell. On the other hand, an exponentially localized tail state close to the mobility edge can have a spread that is of the order of the super cell size, causing an underestimation of the mobility gap. However, when the size of the super cell is increased, the spread of an extended state will increase accordingly, whereas the spread of an exponentially localized state will converge to a finite value. This separates tail and band states, and should lead to convergence of the mobility gap if the cell size is chosen large enough.
Another effect of the finite size is a broadening of bands due to the interaction of semi-localized states with their periodic images, which could shift states out of the gap and therefore smear out the transition between band and tail states. 

\begin{figure}
\begin{center}
\includegraphics[width=1.\textwidth]{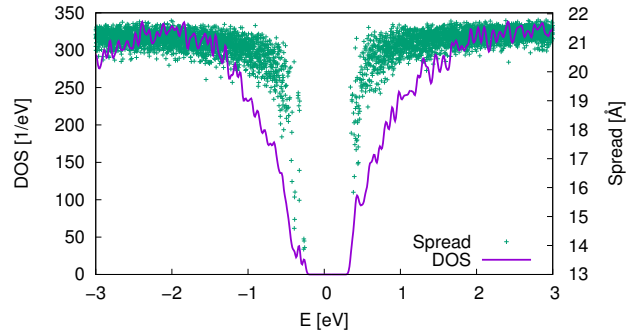}
\caption{DOS and spread of the large system. Each dot represents one wave function.}
\label{spread_512} 
\end{center}
\end{figure}

The distribution of the spread in the large system (\figref{spread_512}) shows indeed the effects of the larger super cell size. The relative difference in the spread of the band states is much smaller, and the tails are more pronounced. Also the mobility gap is more distinct and slightly larger with a value of 1.0 eV (obtained with a threshold of $S=20.7\un{\AA}$), which is however still small as compared to the experimental value of around 1.9 eV \cite{winer:91}. The DOS does not show any mid-gap states, which is in agreement with the fact that the large configuration does not have any dangling bonds. There are however two small peaks at the border of the gap arising from weak bonds, i.e., bonds that show a comparably low value for the ELF between the atoms.

\section{Optical properties}

\subsection{G$_0$W$_0$ calculations}

\begin{figure}
\begin{center}
\includegraphics[width=1.\textwidth]{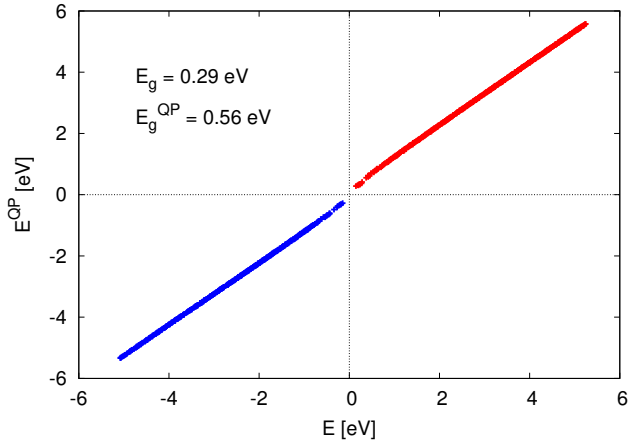}
\caption{Quasi-particle corrected vs. uncorrected electron energies. $E_g$ refers here to the energy difference between the lowest unoccupied and the highest occupied state.}
\label{qp} 
\end{center}
\end{figure}

\figref{qp} shows the quasi-particle corrected electron energies as obtained from the G$_0$W$_0$ calculation for the small a-Si:H structure. The results show that the effect of the quasi-particle corrections consists mainly in a spreading of valence and conduction band by approximately $0.27$ eV, which suggests that the costly G$_0$W$_0$ calculation can be substituted by a simple scissors shift. A scissors shift is a linear shift $E_{v/c}^{QP} = a_{v/c}\cdot E_{v/c}+b_{v/c}$ of both the valence and the conduction band, where $E_{v/c}$ are the uncorrected energies \cite{godby:88}. The respective shifting parameters are obtained by a linear fit of the G$_0$W$_0$ results. The choice of the right set of parameters thereby depends on the energy range of interest. By using different energy ranges for fitting we obtain different parameter sets, which are listed in table \ref{SS}.

\begin{table}
\caption{Scissors shift parameters for a-Si:H obtained from fitting the results of a G$_0$W$_0$ calculation in different energy ranges.}
\begin{tabular}{@{}c|cccc@{}}
\hline
\hline
Fitting range [eV] & $a_v$ & $b_v$ [eV] & $a_c$ & $b_c$ [eV] \\ 
\hline 
$[-1:1]$ & $1.088$ & $-1.097$ & $1.146$ & $-1.228$ \\
$[-2:2]$ & $1.044$ & $-0.858$ & $1.064$ & $-0.665$ \\
$[-3:3]$ & $1.025$ & $-0.762$ & $1.043$ & $-0.509$ \\
\hline
\hline
\end{tabular}
\label{SS}
\end{table}

%\begin{figure}
%\begin{center}
%\includegraphics[width=0.32\textwidth]{fit1.eps}
%\includegraphics[width=0.32\textwidth]{fit2.eps}
%\includegraphics[width=0.32\textwidth]{fit3.eps}
%\caption{Linear fits of the quasi-particle energies using different fitting ranges. Left: -1:1 eV. Center: -2:2 eV. Right: -3:3 eV.}
%\label{fit} 
%\end{center}
%\end{figure}

\subsection{Absorption spectrum}

\begin{figure}
\begin{center}
\includegraphics[width=1.\textwidth]{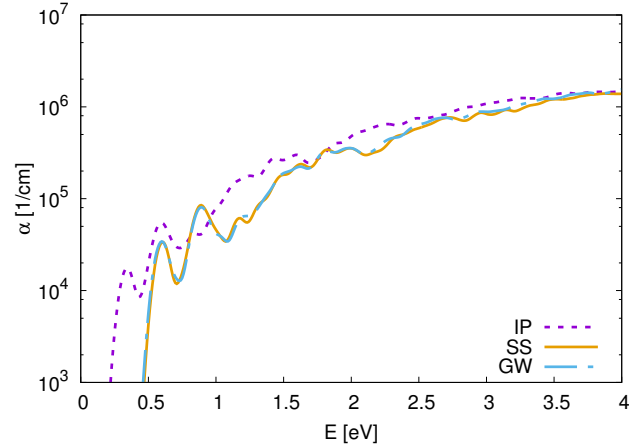}
\caption{Absorption spectrum of the small configuration calculated in independent particle (IP, dotted), GW (dash-dotted), and scissors shift (SS, solid) approximation.}
\label{alpha64} 
\end{center}
\end{figure}

\figref{alpha64} shows the absorption spectrum of the small configuration calculated within the independent-particle (IP) approximation (i.e., with the uncorrected Kohn-Sham energies), the GW approximation, and the scissors-shift (SS) approximation. The IP spectrum shows two sub-gap absorption peaks at 0.34 eV and 0.60 eV. By comparison with the DOS the first peak can be related to absorption processes between two gap states, whereas the second peak arises from absorption processes between a gap state and a tail state. In order to estimate the optical gap we use a Tauc plot \cite{tauc:66}, where the linear regime of $\sqrt{\alpha E}$ is extrapolated and $E_g$ is determined as the intersection of the extrapolated line with the energy axis. This yields a value of $E_g = 0.7\un{eV}$, which is small compared to the experimental value of approximately $1.7\un{eV}$ \cite{cody:81}.

\begin{figure}
\begin{center}
\includegraphics[width=1.\textwidth]{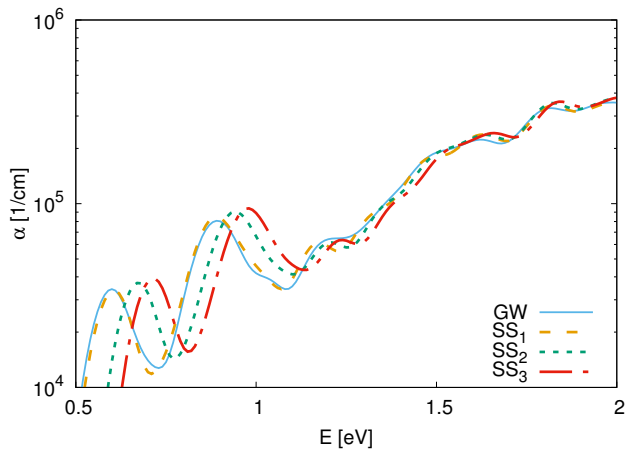}
\caption{Comparison of absorption spectrum of the small configuration calculated in GW approximation (solid) and with different sets of scissors shift parameters (SS; dashed, dotted, and dash-dotted).}
\label{qpvsss} 
\end{center}
\end{figure}

The G$_0$W$_0$ correction modifies the absorption spectrum only in terms of a shift and a slight stretch, which results in a corrected optical gap of $1.0 \un{eV}$. The figure shows that the G$_0$W$_0$ correction can be well approximated by a scissors shift, where the first parameter set in table \ref{SS} was used. These parameters were chosen because they best approximate the G$_0$W$_0$ absorption spectrum, as can be seen in \figref{qpvsss}, where the spectra for all three parameter sets are compared to the G$_0$W$_0$ spectrum. 

\begin{figure}
\begin{center}
\includegraphics[width=1.\textwidth]{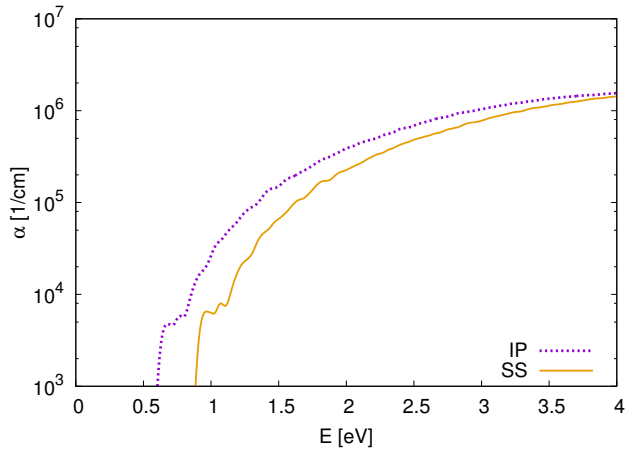}
\caption{Absorption spectrum of the large configuration calculated in IP (dotted) and SS (solid) approximation.}
\label{alpha512} 
\end{center}
\end{figure}

After finding a suitable set of scissors shift parameters for a-Si:H, we use these parameters to calculate a quasi-particle corrected absorption spectrum also for the large configuration, for which a G$_0$W$_0$ calculation would be too costly. The result is shown in \figref{alpha512}, together with the uncorrected spectrum. As compared to the small configuration, the sub-gap absorption decreased significantly. Moreover the optical gap obtained from the Tauc plot increased to $1.0\un{eV}$ in the IP approximation, and to $1.3\un{eV}$ with scissors shift corrections. Whereas the relation between a larger super cell and a decreased sub-gap absorption can be easily understood in terms of the reduced defect density and the reduced spatial overlap of localized states, we do not have a straightforward explanation for the dependence of the optical gap on the super cell size. The results suggest however that finite-size effects do have an influence on $E_g$, and that a closer agreement with the experimental value could be reached with even larger super cells.

\begin{figure}
\begin{center}
\includegraphics[width=\textwidth]{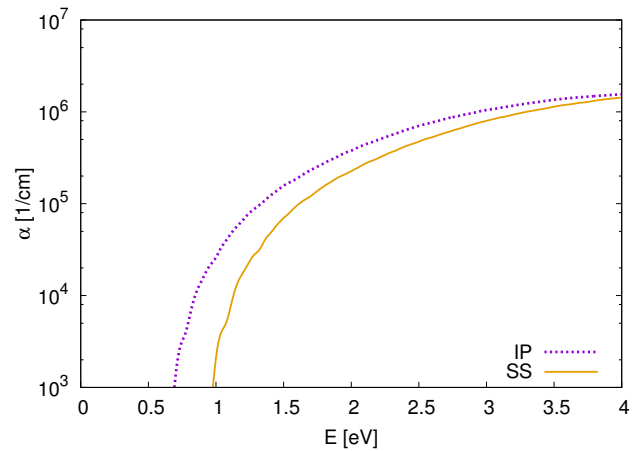}
\caption{Absorption spectrum in IP (dotted) and SS (solid) approximation averaged over 10 large configurations.}
\label{alphaav} 
\end{center}
\end{figure}

The spectra shown so far were obtained from single configurations and have therefore limited physical significance. In order to obtain a physically meaningful absorption spectrum of a-Si:H, i.e., a spectrum that can be compared to experimental spectra, the configurational average has to be taken. For that purpose we calculated spectra for 10 different large configurations and averaged over them. The result is shown in \figref{alphaav}. While sub-gap absorption is still present in the averaged spectrum, the distinct peaks disappeared. This resembles more the experimental findings \cite{tsai:79,cody:81}, even though the contribution of sub-gap absorption is still overestimated. The value of the optical gap is not affected by the averaging.

\section{Conclusions and outlook}
We generated configurations of defective and defect-free a-Si:H consisting of up to 576 atoms, and calculated and subsequently analyzed their structural, electronic, and optical properties. We found that the finite size of the super cell affects both the localization of the electronic states and the absorption spectrum. In particular, a larger super cell improved the values we calculated for the mobility gap and the optical gap, even though a discrepancy of 0.7 to 0.9 eV with the experimental values remains. Furthermore, we successfully performed G$_0$W$_0$ calculations for an a-Si:H configuration with 72 atoms and found that the quasi-particle corrections can be approximated by a scissors shift. This approximation makes calculations for larger -- and thus physically more representative -- configurations possible. The extracted set of scissors shift parameters was used for calculating quasi-particle corrected absorption spectra for all configurations, which improved the optical gap by roughly 0.3 eV. Whether the remaining discrepancy between simulation and experiment is purely a result of finite size effects, and could thus be eliminated by making the super cell large enough, or whether the atomic or electronic structure is not sufficiently well described by our current methods remains an open question and will have to be further investigated.

\begin{acknowledgments}

The authors gratefully acknowledge funding from the European Commission Horizon 2020 research and innovation program under grant agreement No. 676629, support through the COST action MP1406 MultiscaleSolar, as well as the computing time granted on the
supercomputers JURECA \cite{jureca} at J\"ulich Supercomputing Centre and CRESCO \cite{cresco} on the ENEA-GRID infrastructure.

\end{acknowledgments}

\bibliography{czaja}

\end{document}